\documentclass[10pt]{iopart}
\usepackage{iopams,graphicx,color}
\usepackage{url, bm}
\usepackage{tikz}
\usetikzlibrary{calc}
\usetikzlibrary{intersections}

\newcommand{\D}{{\rm d}}
\begin{document}
\title[Relativistic Corrections for Time and Frequency Transfer in Optical Fibres]{Relativistic Corrections for Time and Frequency Transfer in Optical Fibres}
\author{J Ger\v{s}l$^1$, P Delva$^2$ and
P Wolf$^2$}

\address{$^1$ Czech Metrology Institute, Okru\v{z}n\'{i} 31, 63800 Brno, Czech Republic}
\address{$^2$ LNE-SYRTE, Observatoire de Paris, CNRS UMR8630, UPMC, LNE, 61~avenue de l'Observatoire, F-75014 Paris, France}
\ead{jgersl@cmi.cz}
\begin{abstract}
We derive relativistic corrections for one-way and two-way time and frequency
transfer over optical fibres neglecting no terms that exceed 1 ps in time and
$10^{-18}$ in fractional frequency, and estimate their magnitude in typical
fibre links. We also provide estimates of the uncertainties in the evaluation of
the relativistic corrections due to imperfect knowledge of parameters like the
coordinates of the fibre and stations, Earth rotation, or thermal effects of the
fibre index and length. The links between Teddington(UK) and Paris(F) as well as
Braunschweig(D) and Paris(F), that are currently under construction, are studied
as specific examples.
\end{abstract}
\vspace{2pc}

\submitto{\MET}
\maketitle
\ioptwocol

\section{Introduction}

Atomic clocks have been improving rapidly over the past years and are now
reaching stabilities and accuracies of a few parts in $10^{18}$ in fractional
frequency \cite{Hinkley,Bloom}. Applications of such clocks in fundamental
physics, geodesy, navigation etc... require their comparison over large
distances without degrading their performance. At present, no existing satellite
or other long (intercontinental) distance comparison method reaches the required
level of uncertainty. However, over short to medium distances ($10^3$~km)
optical fibre links have demonstrated performance below $10^{-18}$ in frequency
transfer \cite{Predehl,Lopez,Droste} and such fibre networks are therefore one
of the key technologies being developed for the application of the new
generation of atomic clocks in many fields.

The increasing requirements on time/frequency accuracy and stability pushes the
evaluation of phenomena that affect the signal propagation in fibre links,
and the models for their correction, to new levels of uncertainty. One ensemble
of such corrections are related to the evaluation of relativistic effects for a fibre which is moving with given velocity due to the Earth surface motions (rotation, tides) and which is exposed to the Earth gravity field. Systematic relativistic theory of time and frequency transfer
has been worked out e.g. for the case of satellite transfer so far \cite{Wolf1,
Blanchet, Teyssandier1, Teyssandier2}.

In our work we present a systematic relativistic description of signal
propagation in optical fibres and we derive relativistic corrections with an
uncertainty that is sufficient for the new generation of atomic clocks. We
provide expressions for one-way and two-way time and frequency transfer
neglecting no terms that can exceed 1 ps in time and $10^{-18}$ in fractional
frequency. We then estimate the magnitude of these terms and their uncertainty
from imperfect knowledge of the required parameters (fibre and station
coordinates, Earth rotation, fibre index variations,...). Finally, we evaluate
the expressions for time and frequency transfer examples in Europe, in
particular on the PTB-SYRTE and NPL-SYRTE links that are under construction.

In the main part of the text the resulting formulas for relativistic corrections are summarized and the technical details of the derivation can be found in the appendix.


\section{Definitions}

In this section we define some concepts and quantities which will be used for the formulation of relativistic corrections for time and frequency transfer.

First we consider a signal propagating in an optical fibre and an observer who is not moving with respect to the fibre at a certain fibre point. We suppose that a coordinate velocity of the signal as measured in a local coordinate system of this observer, or more precisely a magnitude of the velocity component which is tangent to the fibre, is given as
\begin{equation}
v=\frac{c}{n_{\textrm{eff}}}
\label{m:neff}
\end{equation}
where $n_\textrm{eff}$ is an effective refractive index. The
formula~(\ref{m:neff}) can be understood as definition of the effective refractive
index. Its value should be determined
experimentally for a particular fibre. This value can differ from a value of
refractive index for free propagation in the medium due to interaction of the light
signal with the walls of the fibre. The value of effective refractive index also
depends on temperature of the fibre, stress in the fibre, frequency of the
signal, polarisation of the signal, possibly on fibre bending, etc. For a
waveguide, $n_\textrm{eff} = \beta / \beta_0$ where $\beta$ and $\beta_0$ are
respectively the propagation constants of the wave along the waveguide direction
and in vacuum. We will consider in the following that we can describe the signal
by a light ray. Therefore we do not describe
wave effects such as dispersion, polarisation or interferences explicitly. Nevertheless, waveguide
effects are accounted for by the fact that the effective index can change
with time and along the fiber, and by the fact that the light ray represent one
(transverse) guided mode of the fibre with propagation constant $\beta$. In the following text we drop the index $\textrm{eff}$ and we denote the effective refractive index just $n$. 

Throughout this work we use the Geocentric Celestial Reference System (GCRS).
This coordinate system is centered in the center of mass of the Earth and is
non-rotating with respect to distant stars. The
GCRS coordinates are denoted $(x^0, x^i)$ where $x^0/c=t$ is the Geocentric
Coordinate Time (TCG) and $x^i=(x,y,z)$ are the spatial coordinates of the
system. The small Latin indices go from 1 to 3. 

Components of the metric in the GCRS coordinates are \cite{Soffel}
\begin{eqnarray}
g_{00}&=&-1+\frac{2w}{c^2}-\frac{2w^2}{c^4}+O(c^{-5})\label{m:g00}\\
g_{0i}&=&-\frac{4}{c^3}w_i+O(c^{-5})\label{m:g0i}\\
g_{ij}&=&\delta_{ij}\left(1+\frac{2w}{c^2}\right)+O(c^{-4})\label{m:gij}
\end{eqnarray}
where $w$ and $w_i$ are respectively a scalar and a vector potential
defined in \cite[eq.(20)]{Soffel} with convention $w\geq~\!\!0$.

We can treat the spatial slices of our spacetime given by $t=$ const. as Euclidean spaces, similarly as in Newtonian physics.  The metric of these Euclidean spaces can be defined by components $\delta_{ij}$ in the spatial GCRS coordinates $x^i$. The tangent vectors of these spaces (analogs of vectors in Newtonian physics) we denote in bold and their scalar product given by the Euclidean metric by dot. We denote ${\bf e}_i$ the coordinate basis of the coordinates $x^i$ and thus we have ${\bf e}_i \cdot {\bf e}_j=\delta_{ij}$.

The fibre is considered to be a one dimensional object. Its trajectory in spacetime is therefore parametrised by two parameters - the coordinate time $t$ and a parameter $\lambda$ which has a unique value for a given fibre element and is growing from an initial point of the fibre which we denote $I$ to the final point of the fibre which we denote $F$. The fibre trajectory is given parametrically as
\begin{equation}
x^i=x^i(t,\lambda)\ .
\end{equation}
For a fixed value of $\lambda$ this equation gives a trajectory of a corresponding fibre element. We can define a velocity vector field of the fibre in the GCRS frame and a tangent vector field of the fibre with parameter $\lambda$ as
\begin{equation}
{\bf v}=\frac{\partial x^{i}}{\partial t}{\bf e}_i,\ \ \ {\bf s}_\lambda =\frac{\partial x^i}{\partial\lambda}{\bf e}_i\ .
\label{m:defvs}
\end{equation}

\section{Time transfer}

\begin{figure}
\resizebox{1.0\linewidth}{!} {\begin{tikzpicture}

\tikzstyle{signal} = [thick, color=red]
\tikzstyle{timeline} = [color=black!60]

\path (10,0) node (abs) {space};
\path (0,7) node (ord) {time};

\draw[-latex] (-0.5,0) -- (abs);
\draw[-latex] (0,-0.5) -- (ord);


\path (2.4,0) coordinate (originA);
\path (4.8,0) coordinate (originB);

\path (1.5,6) node[above] (endA) {clock I};
\path (8,6) node[above] (endB) {clock F};

\draw[-latex,timeline] (originA) .. controls +(90:3cm) and +(-90:3cm) .. (endA)
    coordinate[pos=0.15] (tau0)
    coordinate[pos=0.85] (tau2);

\draw[-latex,timeline] (originB) .. controls +(60:3cm) and +(-100:3cm) .. (endB)
    coordinate[pos=0.4] (tau1);


\draw[dashed] let \p1 = (tau0) in (0,\y1) node[left] (t0){$t_0$} -- 
    (\p1) node[below right] (t1l){$\tau_{I0}$};
\draw[dashed] let \p1 = (tau1) in (0,\y1) node[left] (t1){$t_1$} -- 
    (\p1) node[right] (t2l) {$\tau_{F1}$};
\draw[dashed] let \p1 = (tau2) in (0,\y1) node[left] (t2){$t_2$} -- 
    (\p1) node[above right] (t4l){$\tau_{I2}$};


\foreach \e in {tau0,tau1,tau2} {
  \fill[red] (\e) circle(1.5pt);
}


\coordinate (tmp) at ($(tau0)!0.5!(tau1)$);
\draw[-latex, signal] (tau0) .. controls +(50:0.5cm) and +(-150:0.5cm) .. (tmp);
\draw[signal] (tmp) .. controls +(30:0.5cm) and +(-130:0.5cm) ..  (tau1);

\coordinate (tmp) at ($(tau1)!0.5!(tau2)$);
\draw[-latex, signal] (tau1) .. controls +(130:0.5cm) and +(-50:0.5cm) .. (tmp);
\draw[signal] (tmp) .. controls +(130:0.5cm) and +(-50:0.5cm) ..   (tau2);

\node at (2.4,2.7) {$\tau_{I1}$};
\draw [fill] (2.13,2.45) circle(1.5pt);

\end{tikzpicture}}
\caption{The signal leaves observer $I$ at coordinate time $t_0$ corresponding to proper time $\tau_{I0}$ of clock $I$, is
reflected from observer $F$ at coordinate time $t_1$ corresponding to proper time $\tau_{F1}$ of clock $F$ and $\tau_{I1}$ of clock $I$, and finally is
received by the observer $I$ at coordinate time $t_2$ corresponding to proper time $\tau_{I2}$ of clock $I$.
\label{fig:2way}}
\end{figure}
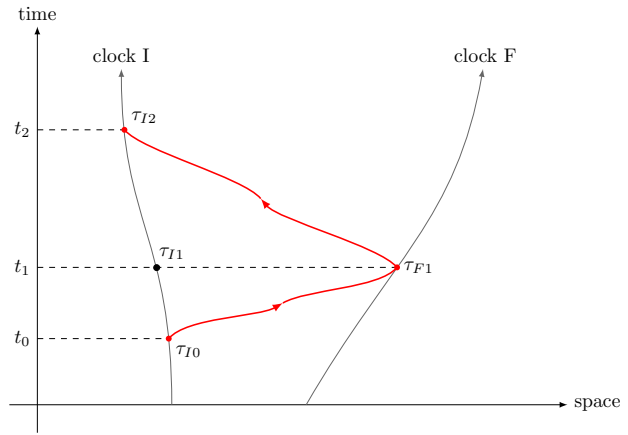

We consider a signal emitted from observer $I$ at coordinate time $t_0$ which
corresponds to proper time $\tau_{I0}$ of clock $I$. The signal is then received by observer
$F$ at coordinate time $t_1$ which corresponds to proper time $\tau_{F1}$ of
clock $F$ (see figure~\ref{fig:2way}). The "pseudo-time-of-flight"
$\tau_{F1} - \tau_{I0}$ is obtained from measurements. Then a signal is sent from observer $F$
at coordinate time $t_1$ and received by observer $I$ at coordinate time $t_2$
which corresponds to proper time $\tau_{I2}$ of clock $I$. This signal can be
either the same signal reflected or another signal which is synchronously sent
when the one-way signal is received by observer $F$. We call this set-up the
$\Lambda$-configuration. The "pseudo-time-of-flight" $\tau_{I2} -
\tau_{F1}$ is also obtained from measurements. Using the coordinate time synchronisation
convention we define $\tau_{I1} = \tau_I (t_1)$.

We denote $\Delta t_+ = t_1-t_0$ the coordinate time of signal propagation from $I$ to $F$ and $\Delta t_- =  t_2-t_1$ the coordinate time of signal propagation from $F$ to $I$.

It will be convenient to introduce a special choice of the fibre parameter $\lambda$ which is given by a rest length of the fibre from its initial point $I$ at time $t_1$. We denote this parameter $l$ (see \ref{App2} and formula (\ref{lt}) for a precise definition). The range of this parameter is $l\in[0,L]$ with $L$ being the total rest length of the fibre at the time $t_1$.

For the coordinate propagation times we obtain the following formula up to $c^{-3}$ order (see \ref{App} for derivation) which gives all the terms larger than 1~ps
\begin{eqnarray}
\Delta t_\pm&=&\frac{1}{c}\int\limits_0^{L}n\ \D l\pm\frac{1}{c^2}\int\limits_0^{L}{\bf v}\cdot{\bf s}_l\ \D l\label{m:dtp1}\\
&&+\frac{1}{c^3}\int\limits_0^{L}n\left(w+{v^2}/{2}\right)\D l \nonumber
\end{eqnarray}
where $v^2={\bf v}\cdot{\bf v}$ and all quantities in this formula are evaluated at $t=t_1$.\footnote{If needed the integrals can be expressed in terms of Euclidean length parameter $l_E$ instead of $l$ using the transformation formula (\ref{EuclidRest}).} 

For a particular case of a 1000~km long optical fibre with refractive index 1.5 we obtain the following numerical estimates of the terms in (\ref{m:dtp1}). The leading Newtonian term (the $c^{-1}$ term) has a
value of 5~ms. If the fibre is located along the equator the value of $c^{-2}$ term is $\pm 5$~ns. This term corresponds to the Sagnac correction as we will see.
The $c^{-3}$ term gives 3~ps at surface of the Earth. This term is a part of Shapiro correction. Another part appears if we use the Euclidean length instead of the rest length as the fibre parameter (see \ref{App} for further details). Higher order terms in $1/c$ are negligible for the 1~ps accuracy level. E.g. the $c^{-4}$ term is of order of $10^{-23}$~s per one meter of the fibre length.

Analogous corrections appear also for the satellite time transfer (see e.g. \cite{Blanchet}). However, in case of fibres, where the signal propagation is not geodesic, the state of the fibre given by its position, velocity and refractive index enters the formulas. This is the difference compared to the satellite transfer where only the state of the emitter and receiver appears.

The full formula (\ref{m:dtp1}) would be used in case of one-way time transfer. However, in practise it is difficult to evaluate the leading term of this formula with sufficient accuracy. Therefore the two-way time transfer is used which compensates this term. 

Using the coordinate synchronisation convention, the desynchronisation of the clocks is defined as difference $\tau_{F1}-\tau_{I1}$. In case of the two-way time transfer it can be expressed as
\begin{equation}
\tau_{F1}-\tau_{I1}=\frac{1}{2}(\tau_{F1}-\tau_{I0}+\tau_{F1}-\tau_{I2})+\frac{1}{2}(\Delta\tau_- -\Delta\tau_+)
\label{m:desyn}
\end{equation}
where $\Delta\tau_+ =\tau_{I1}-\tau_{I0}$ and $\Delta\tau_-
=\tau_{I2}-\tau_{I1}$. The "pseudo-time-of-flights" $\tau_{F1}-\tau_{I0}$ and
$\tau_{I2}-\tau_{F1}$ in (\ref{m:desyn}) are measured and the difference $\Delta\tau_- -\Delta\tau_+$ needs to be computed. The computed term can be approximated as 
\begin{equation}
\Delta\tau_- -\Delta\tau_+\approx \Delta t_- -\Delta t_+\ . \label{dtauapp}
\end{equation}
Error of this approximation is several orders below the required 1~ps level for a fibre at the Earth surface.
Using (\ref{m:dtp1}) we obtain
\begin{equation}
\frac{1}{2}(\Delta t_- -\Delta t_+) = -\frac{1}{c^2}\int\limits_0^{L}{\bf v}\cdot{\bf s}_l\ \D l\ .\label{m:twtt2}
\end{equation}
The term ${\bf s}_l\D l=\D x^i{\bf e}_i$ does not depend on choice of the fibre parameter ($l$ in this case) and therefore the integral in (\ref{m:twtt2}) can be easily transformed if any other parameter (e.g. the Euclidean length) is more convenient.

The main part of the velocity in the formula (\ref{m:twtt2}) and in the second term of formula~(\ref{m:dtp1}) comes from the Earth rotation.  
To interpret this term we introduce a new coordinate frame which rotates rigidly with respect to the Euclidean frame of the GCRS system with angular velocity vector $\bomega$. We suppose that this rotating frame follows the rotation of the Earth surface such that the residual velocity of the fibre in this rotating frame is minimized. We refer to this frame as to a co-rotating frame.   
We denote the position vector of a fibre point by ${\bf x}$ and the velocity vector of a fibre point in the co-rotating frame by
${\bf v}_R$. We have
\begin{equation}
{\bf v}={\bomega}\times{\bf x}+{\bf v}_R\ .
\end{equation}
For the scalar product in (\ref{m:twtt2}) we get
\begin{eqnarray}
{\bf v}\cdot {\bf s}_l &=& ({\bomega}\times{\bf x})\cdot {\bf s}_l+{\bf v}_R\cdot {\bf s}_l\\
&=& {\bomega}\cdot({\bf x}\times {\bf s}_l)+{\bf v}_R\cdot {\bf s}_l\ .\nonumber
\end{eqnarray}
If we write $\bomega=\omega {\bf o}$ where ${\bf o}\cdot{\bf o}=1$, i.e. $\bf o$ is a unit vector in direction of $\bomega$ we can define a quantity
\begin{equation}
A=\frac{1}{2}\int\limits_{0}^{L}{\bf o}\cdot({\bf x}\times {\bf s}_l)\ \D l\ .\label{A}
\end{equation}
This quantity is the Sagnac area of the fibre which for certain simple fibre
paths can be understood as area of a surface which is lying in a plane perpendicular to $\bomega$
between a projection of the fibre into this plane and lines connecting
endpoints of this projection with the rotation axis. This area can
be positive or negative depending on whether the parameter of the fibre ($l$ in
this case) goes along or against the direction of Earth rotation. Using the Sagnac area $A$ we can write
\begin{equation}
\int\limits_0^{L}\!{\bf v}\cdot{\bf s}_l\ \D l=2\omega A+\int\limits_0^L\!{\bf v}_R\cdot {\bf s}_l\ \D l\ .\label{SagVec}
\end{equation}
In this formula the first term is usually much larger than the second one. We can estimate the second term if the velocity ${\bf v}_R$ is caused by the Earth tides. The Earth tides are oscillatory deformations of the Earth body with period of 12~h and amplitude of vertical Earth surface motion of approx. 30~cm leading to maximal vertical velocity of the Earth surface of approx. 0.05~mm/s. The effect would be maximal for a fibre path following a meridian near $45^\circ$ latitude. The second term of (\ref{SagVec}) then gives 0.3~fs contribution to the correction (\ref{m:twtt2}) per 1000~km of the fibre in this maximizing example. 

The phenomena affecting the time transfer are summarized in table~\ref{tabTT1}.
\begin{table}
\begin{center}
{
\small
\begin{tabular}{l c}
\hline
Effect & Contribution\\
& per 1000~km\\
\hline
Length and refractive index of the fibre&\\
(Newtonian term; 1-way only) & 5 ms\\
Velocity of the fibre due to&\\
the Earth rotation&5~ns\\
Velocity of the fibre due to&\\
the Earth tides &0.3~fs\\
Gravitational plus centrifugal potential&\\
on the Earth surface (1-way only) & 3~ps\\
\hline
\end{tabular}
\caption{\label{tabTT1} Effects influencing the time transfer in optical fibres and sizes of the corresponding contributions to the 1-way and 2-way time transfer formulas. All contributions are calculated for 1000 km of the fibre length. The values of the second and the third contribution (Sagnac correction) are examples for specific fibre positions where the effect is maximized (see the text for details).}
}
\end{center}
\end{table}

\section{Frequency transfer}

\begin{figure}
\resizebox{1.0\linewidth}{!} {\begin{tikzpicture}

\tikzstyle{signal} = [thick, color=red]
\tikzstyle{timeline} = [color=black!60]

\path (10,0) node (abs) {space};
\path (0,7) node (ord) {time};

\draw[-latex] (-0.5,0) -- (abs);
\draw[-latex] (0,-0.5) -- (ord);


\path (2.4,0) coordinate (originA);
\path (4.8,0) coordinate (originB);

\path (1.5,6) node[above] (endA) {emitter};
\path (8,6) node[above] (endB) {receiver};

\draw[-latex,timeline] (originA) .. controls +(90:3cm) and +(-90:3cm) .. (endA)
    coordinate[pos=0.15] (taue) 
    coordinate[pos=0.45] (taue2);

\draw[-latex,timeline] (originB) .. controls +(60:3cm) and +(-100:3cm) .. (endB)
    coordinate[pos=0.4] (taur)
    coordinate[pos=0.7] (taur2);




\foreach \e in {taue,taue2,taur,taur2} {
  \fill[red] (\e) circle(1.5pt);
}


\coordinate (tmp) at ($(taue)!0.5!(taur)$);
\draw[-latex, signal] (taue) .. controls +(50:0.5cm) and +(-150:0.5cm) ..  (tmp);
\draw[signal] (tmp) .. controls +(30:0.5cm) and +(-130:0.5cm) .. (taur);
\path[signal] (tmp) node[below] (lf1) {$S$};

\coordinate (tmp) at ($(taue2)!0.5!(taur2)$);
\draw[-latex, signal] (taue2) .. controls +(50:0.5cm) and +(-150:0.5cm) .. (tmp);
\draw[signal] (tmp) .. controls +(30:0.5cm) and +(-130:0.5cm) .. (taur2);
\node at (4.7,3.05) {\color{red}$S + \D S$};

\node[left] at (taue) {$\tau_e$};
\node[left] at (taue2) {$\tau_e+\D \tau_e$};
\node[right] at (taur) {$\tau_r$};
\node[right] at (taur2) {$\tau_r + \D \tau_r$};

\end{tikzpicture}}
\caption{Two clocks are measuring
   proper time along their trajectory. One signal with phase $S$ is emitted at proper time $\tau_e$, and another one with phase $S+\D S$ at time
   $\tau_e + \D \tau_e$. They are received respectively at time
   $\tau_r$ and $\tau_r+\D \tau_r$. The proper frequency measured by the
   emitter/receiver is respectively:
$ \nu_{e/r} = \frac{1}{2\pi} \frac{\D S}{\D \tau_{e/r}}$.\label{fig:2wayfreq}}
\end{figure}
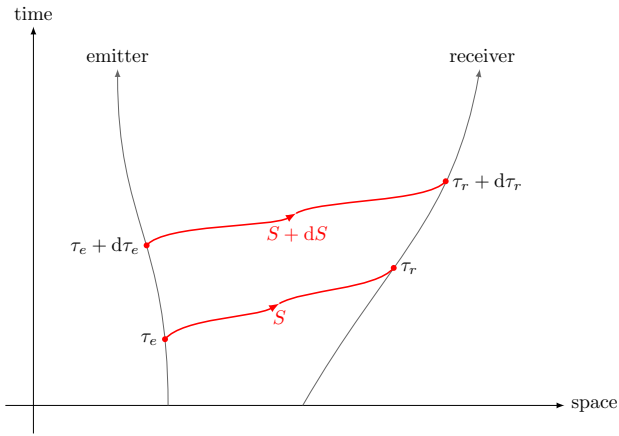

For one-way frequency transfer we consider that a signal with proper frequency
$\nu_e$ is emitted from one observer and the same signal is received by a second
observer with proper frequency $\nu_r$. We suppose that a phase of the signal
which is observed by the emitting observer at its proper time $\tau_e$ is
observed by the receiving observer at its proper time $\tau_r$. This is
illustrated on figure~\ref{fig:2wayfreq}.

The time $\tau_r$ is uniquely given by the time $\tau_e$ so we can define a
function $\tau_r(\tau_e)$. The ratio of the proper frequencies is then given as
\begin{equation}
\frac{\nu_r}{\nu_e}=\left(\frac{\D \tau_r}{\D \tau_e}\right)^{-1}=\left(\frac{\D t_e}{\D \tau_e}\right)^{-1}\frac{\D t_r}{\D \tau_r}\left(\frac{\D t_r}{\D t_e}\right)^{-1}.\label{m:fratio}
\end{equation}

The derivatives ${\D t_e}/{\D \tau_e}$ and ${\D t_r}/{\D \tau_r}$ do not depend
on the fibre itself but just on the state of the emitting and receiving
observers. They contain the gravitational red shift and second order Doppler effect corrections. The same terms appear also for different ways of frequency transfer
such as the satellite transfer (see e.g. \cite{Blanchet}). These terms are investigated in detail
in~\cite{Wolf1} and it is not the purpose of this paper to study
them. For a relative accuracy of the frequency transfer of $10^{-18}$, these terms contain the Newtonian
gravitational potential of the Earth and of external masses, as well as a
correction for the non-geodesic barycentric motion of the Earth. 
Recently these terms have been determined with accuracy of $10^{-17}$ \cite{Calonico}. 
At present, a combination of terrestrial data sets with up-to-date satellite gravity field models allows the calculation of these terms with accuracies of a few parts in $10^{18}$, provided that high-resolution and high-quality terrestrial and satellite data are utilised \cite{Denker}. Related research is tackled in
the frame of the project ITOC (International Timescales with Optical
Clocks)~\cite{ITOC}.  On the other hand, the frequency comparison can be used
to measure directly the geopotential difference between two clocks, a technique
called chronometric leveling~\cite{petit_brumberg}. A large-scale demonstration
of chronometric leveling is also one of the task of the project ITOC.

In this paper we focus on how the frequency transfer is affected by processes in the fibre itself. These effects are contained in the $\D t_r/\D t_e$ term of (\ref{m:fratio}). This term includes e.g. influences of changing refractive index, changing fibre length due to thermal expansion or changing Sagnac area due to fibre motion. 

We denote $T(t,l)$ a temperature of the fibre as a function of time and location and $\alpha$ the linear thermal expansion coefficient of the fibre. Again we consider a signal sent from $I$ at time $t_0$, reflected in $F$ at time $t_1$ and received back by $I$ at time $t_2$. The $\D t_r/\D t_e$ term is then given by the following formula up to $c^{-2}$ order (see the \ref{App} for derivation)
\begin{eqnarray}
\frac{\D t_{r\pm}}{\D t_e}&=&1+\frac{1}{c}\int\limits _0^{L}\left(\frac{\partial n}{\partial t}+n\alpha\frac{\partial T}{\partial t}\right) \D l \label{m:dtdti2}\\
&\pm&\frac{1}{c^2}\int\limits_0^{L}\frac{\partial {\bf v}\cdot{\bf s}_l}{\partial t}\ \D l\nonumber
\end{eqnarray}
where $+$ sign relates to propagation from $I$ (emitter) to $F$ (receiver) and $-$ sign to propagation from $F$ (emitter) to $I$ (receiver). All quantities in the formula are evaluated at the time $t=t_1$. The formula includes all terms larger than $10^{-18}$. Order of the $c^{-3}$ term is estimated to be
$10^{-28}$ per one meter of the fibre length due to thermal effects. The $c^{-1}$ term of (\ref{m:dtdti2}) corresponds to the first order Doppler effect and the $c^{-2}$ term is a time derivative of the Sagnac correction. Analogous terms can be found also in the satellite frequency transfer \cite{Blanchet}, however, some differences occur. E.g. the Doppler shift for fibres does not depend on velocities of the emitter and receiver but on expansion and change of the refractive index of the fibre.

Now we look to the $c^{-1}$ term of (\ref{m:dtdti2}) in more detail and we estimate its value. The change of the refractive index with time can be caused by various phenomena. For our estimation we consider a change of $n$ due to temperature variation, i.e. $\partial n/\partial t =(\partial n/\partial T).(\partial T/\partial t)$. 
The values of linear thermal expansion coefficient and temperature
derivative of refractive index, based on~\cite{Cohen}, can be
estimated as 
\begin{equation}
\alpha = 8\times10^{-7}\ {\rm K}^{-1},\ \ \ \frac{\partial n}{\partial T}=10^{-5}\ {\rm K}^{-1}\ .
\label{alpha}
\end{equation}
The time derivative of the temperature can
be estimated based on the experiment described in~\cite{Ebenhag}. Its maximal value in this
experiment was
\begin{equation}
\frac{\partial T}{\partial t}\approx4\times10^{-6}\ {\rm K/s}\ .
\label{dTdt}
\end{equation}
If we consider a 1000~km long fibre with $n=1.5$ we obtain a value of $2\times10^{-13}$ for the $c^{-1}$ term of (\ref{m:dtdti2}). 

In case of one-way frequency transfer the full formula (\ref{m:dtdti2}) is needed, however, in practise it is difficult to determine the $c^{-1}$ term with a sufficient accuracy. Therefore it needs to be actively compensated or suppressed by means of two-way frequency transfer. The $c^{-2}$ term of (\ref{m:dtdti2}) appears also in the two-way transfer and it is discussed in more detail below.

During two-way frequency transfer the observer $I$ emits a signal with proper
frequency $\nu_{I0}$. The signal is received at observer $F$ and
immediately transponded back to the observer $I$ where a proper frequency
$\nu_{I2}$ is measured\footnote{A generalization of the formula in the case
there is a delay between the reception and the re-emission at observer $F$ is
possible.}. The goal is to express the frequency $\nu_{F1}$ which should be
observed by $F$ at the time of reception with use of the measured frequencies
$\nu_{I0}$ and $\nu_{I2}$ and a computed correction. We express a ratio
$\nu_{I2}/\nu_{F1}$ as
\begin{equation}
\frac{\nu_{I2}}{\nu_{F1}}=\frac{1}{2}\frac{\nu_{I2}}{\nu_{I0}}+\Delta +\frac{1}{2}
\label{m:2wayf}
\end{equation}
where the first term on the right hand side contains only the measured
quantities and the correction $\Delta$ needs to be computed. For this correction we obtain (see the \ref{App} for derivation)
%
\begin{equation}
\Delta = \delta_{I1} - \delta_{F1}
+\frac{1}{c^2}\int\limits_0^{L}\frac{\partial{\bf v}\cdot{\bf s}_l}{\partial t}\ \D l\label{m:delta2}
\end{equation}
where the index $I1$ or $F1$ means the quantity is evaluated at the time $t=t_1$
in the initial or final end of the fibre respectively and we introduced
$\delta \ll 1$ such that $\D t/ \D \tau = 1+\delta$. For the required accuracy of $10^{-18}$ the $\delta$-terms in this equation depend only on the scalar potential and velocity at the end-points of
the fibre and are studied in detail in~\cite{Wolf1}. In our case of spacetime with the GCRS metric we obtain
\begin{equation}
\delta \equiv \frac{\D t}{\D \tau}-1=\frac{1}{c^2}\left(w+\frac{v^2}{2}\right)+O(c^{-4})\ .
\label{potRatio2}
\end{equation}

The integral term of (\ref{m:delta2}) is evaluated at the time $t=t_1$ too. This term can be expressed as
\begin{equation}
\int\limits_0^{L}\frac{\partial{\bf v}\cdot{\bf s}_l}{\partial t}\ \D l=\int\limits_0^{L}{\bf a}\cdot{\bf s}_l\ \D l+\frac{1}{2}(v^2_{F1}-v^2_{I1})
\label{dsagnacdt}
\end{equation}
where ${\bf a}=\partial{\bf v}/\partial t=(\partial^2x^i\!/\partial t^2){\bf e}_i$ is acceleration of points of the fibre in the GCRS frame and the second term was obtained using
\[
{\bf v}\cdot \frac{\partial{\bf s}_l}{\partial t}=\delta_{ij}\frac{\partial x^i}{\partial t}\frac{\partial^2 x^j}{\partial t\partial l}=\frac{1}{2}\frac{\partial}{\partial l}\left(\delta_{ij}\frac{\partial x^i}{\partial t}\frac{\partial x^j}{\partial t}\right)=\frac{1}{2}\frac{\partial v^2}{\partial l}.
\]

Inserting (\ref{potRatio2}) and (\ref{dsagnacdt}) into (\ref{m:delta2}) we get another expression
for $\Delta$
\begin{equation}
\Delta = \frac{1}{c^2}\left(w_{I1}-w_{F1}\right)+\frac{1}{c^2}\int\limits_0^{L}{\bf a}\cdot{\bf s}_l\ \D l\ .
\label{delta3}
\end{equation}
We can see that the terms depending on velocity of the end-points cancel.

The largest contribution to the acceleration ${\bf a}$ is given by a centrifugal acceleration due to the Earth rotation. To see the centrifugal acceleration explicitly we can express ${\bf a}$ in terms of quantities related to the frame co-rotating together with Earth as follows
\begin{equation}
{\bf a}=\bomega\times(\bomega\times{\bf x})+2\bomega\times{\bf v}_R+\bvarepsilon\times{\bf x}+{\bf a}_R
\label{arot}
\end{equation}
where $\bomega$ is the angular velocity of the Earth surface, ${\bf v}_R$ is the velocity of the fibre points in the co-rotating frame, $\bvarepsilon$ is angular acceleration of the Earth surface and ${\bf a}_R$ is acceleration of the fibre points in the co-rotating frame. In terms of co-rotating coordinates $x_R^i$ and the corresponding basis ${\bf e}_{i}^R$ the quantities can be expressed as ${\bf x}=x_R^i {\bf e}_{i}^R$, ${\bf v}_R=(\partial x^i_R/\partial t){\bf e}_{i}^R$, ${\bf a}_R=(\partial^2x^i_R\!/\partial t^2){\bf e}_i^R$, $\bomega=\omega^i_R{\bf e}_{i}^R$ and $\bvarepsilon=\varepsilon_R^i{\bf e}_{i}^R$ with $\varepsilon_R^i\equiv(\partial\omega^i_R/\partial t)$. The first term in (\ref{arot}) is the centrifugal acceleration which is followed by the Coriolis and Euler acceleration.
Using the fact that 
\begin{eqnarray}
(\bomega\times(\bomega\times{\bf x}))\cdot{\bf s}_l&=&-(\bomega\times{\bf x})\cdot(\bomega\times{\bf s}_l)\\
&=&-\frac{1}{2}\frac{\partial(\bomega\times{\bf x})^2}{\partial l}\nonumber
\end{eqnarray}
we can integrate the centrifugal term explicitly and express it in terms of centrifugal potential $(\bomega\times{\bf x})^2/2=(\omega R\sin\theta)^2/2$ with $R=||{\bf x}||$ and $\theta$ being an angle between the rotation axis and the position vector of the fibre end-point $\bf x$. Inserting (\ref{arot}) to (\ref{delta3}) then gives
\begin{eqnarray}
\Delta &=& \frac{1}{c^2}\left(w_{I1}-w_{F1}\right)+\frac{1}{c^2}\left[\frac{(\bomega\times{\bf x}_I)^2}{2}-\frac{(\bomega\times{\bf x}_F)^2}{2}\right]\nonumber\\
&+&\frac{1}{c^2}\int\limits_0^{L}({\bf a}_R+2\bomega\times{\bf v}_R+\bvarepsilon\times{\bf x})\cdot{\bf s}_l\ \D l \label{inta}
\end{eqnarray}
where again all quantities are expressed at $t=t_1$.

To obtain the best accuracy for the endpoint terms of (\ref{inta}) given by sum of the scalar gravitational potential $w$ and the centrifugal potential, the two contributions should not be computed separately but the procedure described in \cite{Wolf1} should be followed.

Contributions to the integral term of (\ref{inta}) are given e.g. by Earth tides or by variations in the Earth rotation. Considering the Earth tides with period of 12~h and amplitude of vertical Earth surface motion of approx. 30~cm we obtain maximal vertical velocity of the Earth surface of approx. 0.05~mm/s as we already mentioned and maximal vertical acceleration of approx. $6\times 10^{-9}\ {\rm ms}^{-2}$. The maximal contribution of the Earth tides is given by the Coriolis term in (\ref{inta}) for a fibre path following the equator. Its value in this maximizing example is $8\times10^{-20}$ per 1000~km of the fibre. The maximal contribution of the ${\bf a}_R$ term in (\ref{inta}) due to the Earth tides is around $3\times 10^{-20}$ per 1000~km of a fibre following a meridian near $45^{\circ}$ latitude.

The angular velocity vector $\bomega$ precesses around certain fixed axis in the co-rotating frame with one day period and with evolving amplitude. We choose the basis of the co-rotating frame ${\bf e}_R^i$ such that ${\bf e}_R^z$ points in direction of this axis. The angular acceleration components $\varepsilon_R^x, \varepsilon_R^y$ therefore oscillate with one day period and maximal amplitude of approx. $5\times 10^{-16}\ {\rm s}^{-2}$ mostly due to the changing direction of $\bomega$. This value was estimated based on known evolution of the Earth angular velocity vector in Terrestrial Reference Frame. The component $\varepsilon_R^z$ is several orders below the $x,y$ - components corresponding to the fact that the effect of changing magnitude of $\bomega$ is much smaller than the effect of changing direction. The largest contribution of the Euler term in (\ref{inta}) would be observed for a fibre path following the meridian perpendicular to the instant direction of $\bvarepsilon$ (neglecting the $\varepsilon_R^z$ component). In this maximizing example we obtain the contribution of $4\times 10^{-20}$ per 1000~km of the fibre.

The phenomena contributing to the frequency transfer corrections are summarized in table \ref{tabFT1}.
\begin{table}
\begin{center}
{
\small
\begin{tabular}{l c}
\hline
Effect & Correction\\
\hline
Difference of gravitational plus&\\
centrifugal potential at endpoints&$>10^{-18}$\\
Variations in length and refractive index&\\
due to temperature changes (1-way only)&$\sim 10^{-13}$\\
Coriolis acceleration of the fibre&\\
due to velocity of the Earth tides& $8\times 10^{-20}$\\
Euler acceleration of the fibre due to&\\
angular acceleration of the Earth rotation&$4\times 10^{-20}$\\
Acceleration of the Earth tides & $3\times 10^{-20}$\\
\hline
\end{tabular}
\caption{\label{tabFT1} Effects influencing the frequency transfer in optical fibres and sizes of the corresponding corrections in 1-way and 2-way frequency transfer formulas. The first correction depends on fibre endpoints only. The remaining corrections depend on processes in the whole fibre and they are calculated per 1000~km of the fibre length. Values for the last three corrections are examples for specific fibre positions where the effect is maximized (see the text for details).} 
}
\end{center}
\end{table}

\section{Required uncertainty of the input parameters}

In this section we discuss what is the required uncertainty of the input parameters entering the formulas for time and frequency transfer in order to keep the uncertainty of the resulting corrections below the 1~ps target for time transfer and $10^{-18}$ target for frequency transfer.

\subsection{Time transfer}

The required uncertainty of the length of the fibre in case of one-way time transfer would be 0.2~mm to achieve 1~ps uncertainty of the propagation time. For 1000~km long fibre the required uncertainty of the effective refractive index would be $3\times 10^{-10}$. Since these uncertainties are difficult to achieve in practice, e.g. because of unknown temperature variations, the two-way transfer is used. Therefore we focus to the evaluation of the two-way time transfer formula (\ref{m:twtt2}) only.
         
For evaluation of (\ref{m:twtt2}) we use its form given by (\ref{SagVec}). First we investigate uncertainty of the Sagnac term due to uncertain position of the fibre and we look for a requirement for the position uncertainty leading to the Sagnac term within the 1~ps uncertainty limit. We consider that the fibre position at certain coordinate time is shifted by a vector $\bxi(l)$, i.e. the new position vector is ${\bf x}(l)+\bxi(l)$. The corresponding change of the Sagnac term can be obtained by inserting this new position vector to the formula (\ref{A}) where ${\bf s}_l=\D{\bf x}/\D l$. Expanding this formula and integrating by parts we obtain
\begin{eqnarray}
2\omega\Delta A&=& \bomega\cdot({\bf x}_F\times\bxi_F)-\bomega\cdot({\bf x}_I\times\bxi_I)\label{DA}\\
&+&\!2\!\int\limits_0^{L}\!\!\bomega\cdot\left(\bxi\times({\bf s}_l+\frac{1}{2}\frac{\D\bxi}{\D l})\right)\D l\nonumber
\end{eqnarray} 
where ${\bf x}_I,\bxi_I$ are the position vector and shift vector at the initial end of the fibre and similarly  ${\bf x}_F,\bxi_F$ are the quantities at the final end. The change of the Sagnac term in the time transfer correction (\ref{m:twtt2}) is given as $2\omega\Delta A/c^2$. 

First we look to the contribution of the endpoint terms of the formula
(\ref{DA}). These terms correspond to a change of the Sagnac area given by a shift of the lines connecting the endpoints of the fibre with the Earth center and they are maximized when the endpoints of the fibre lie at the equator and when the shift vectors point along the equator. In this case we have e.g. $\bomega\cdot({\bf x}_F\times\bxi_F)=\omega R_E \xi_F$ with $R_E$
being the Earth radius and $\xi_F=||\bxi_F||$ is the Euclidean magnitude of $\bxi_F$. We get $\omega R_E
\xi_F/c^2=1{\rm ps}$ for $\xi_F\approx 200\ {\rm m}$, i.e. a shift of an endpoint
of the fibre by not more than approx. 200~m can cause a shift in time transfer by not more than 1~ps. 

Next we look to the contribution of the integral term of (\ref{DA}). This term corresponds to the Sagnac area in between the original and shifted fibre paths. If we assume that the shift $\bxi$ does not change the rest length of the fibre, we can derive the following upper bound 
\begin{equation}
\label{err_sagnac}
\left|2\!\int\limits_0^{L}\!\!\bomega\cdot\left(\bxi\times({\bf s}_l+\frac{1}{2}\frac{\D\bxi}{\D l})\right)\D l\right|\leq 2\omega\bar{\xi}L
\label{dAbound}
\end{equation}
where $\bar{\xi}$ is an average shift magnitude defined as
\begin{equation}
\bar{\xi}=\frac{1}{L}\int\limits_0^{L}||\bxi||\ \D l\ .
\end{equation}
The corresponding upper bound for change in the correction (\ref{m:twtt2}) therefore is $2\omega\bar{\xi}L/c^2$. It equals to 1~ps for $\bar{\xi}\approx 600\ {\rm km}^2\times L^{-1}$. For an example of 1000~km long fibre it means that average uncertainty of fibre position better than approx. 600~m is sufficient for 1~ps uncertainty in time transfer.

Including a fibre expansion into $\bxi(l)$ adds a minor correction to the bound (\ref{dAbound}) for expectable fibre length uncertainties.

Next we check the effect of changing $\omega$ for the time transfer. The condition $2\Delta\omega A/c^2=$ 1~ps for an equatorial fibre where $A$ is maximized leads to $\Delta\omega/\omega \approx 200\ {\rm m}\times L^{-1}$. For a 1000~km long fibre it corresponds to a relative uncertainty of order of $10^{-4}$ in $\omega$. The fluctuations of Earth angular velocity are much smaller than this so the Earth angular velocity is not an issue from the uncertainty point of view.

The second term of (\ref{SagVec}) can be expressed as a product $L\bar{v}_{Rt}$ where we defined an average tangent velocity of the fibre in the co-rotating frame as $\bar{v}_{Rt}\equiv\frac{1}{L}\int_0^L\!{\bf v}_R\cdot {\bf s}_l\ \D l$. The 1~ps change in time transfer correction due to a variation in $\bar{v}_{Rt}$ occurs if $L\Delta\bar{v}_{Rt}/c^2=1\ {\rm ps}$, i.e. the maximal allowed uncertainty of $\bar{v}_{Rt}$ for 1~ps time transfer is $\Delta\bar{v}_{Rt}=9\times 10^4\ {\rm m}^2/{\rm s}\times L^{-1}$, which for 1000~km long fibre gives 0.09~m/s. If the fibre is fixed to the Earth surface the magnitude of ${\bf v}_R$ itself is usually much smaller than this. E.g. the maximal velocity caused by the Earth tides is around 0.05~mm/s as already mentioned.

A variation of $L$ in the product $L\bar{v}_{Rt}$ leads to the 1~ps shift in time transfer if $\Delta L\bar{v}_{Rt}/c^2=1\ {\rm ps}$. For expectable values of $\bar{v}_{Rt}$ this leads to values of $\Delta L$ which are much larger than the usual uncertainty of length measurement. Therefore the uncertainty of the fibre length is not an issue in this term.

The input parameters for time transfer and their required uncertainties are summarized in table \ref{tabTT2}.

%
\begin{table}
\begin{center}
{
\small
\begin{tabular}{l c}
\hline
Parameter & Uncertainty\\
\hline
Fibre length (1-way only)& 0.2~mm\\
Refractive index (1-way only)& $3\times 10^{-10}$\\
Fibre endpoints position& 200~m\\
Fibre inner points position& 600~m\\
Fibre velocity in co-rotating frame& 9~cm/s\\
Earth angular velocity&$\sim 0.01$ \% (relative) \\
Gravitational plus centrifugal& \\
potential (1-way only)&$\sim 30$~\% (relative)\\
\hline
\end{tabular}
\caption{\label{tabTT2} Input parameters and their maximal uncertainties sufficient for 1~ps uncertainty in time transfer. The values were obtained for situations where the sensitivity of a correction to a parameter is maximized and they are calculated for 1000~km long fibre (see the text for further details and scaling of the uncertainties with the fibre length).}
}
\end{center}
\end{table}

\subsection{Frequency transfer}

The required uncertainty of time derivative of temperature of the fibre would be $3\times 10^{-11}$~K/s to achieve the $10^{-18}$ uncertainty in one-way frequency transfer for 1000~km long fibre. Since this is difficult to achieve in practise the two-way transfer is used. Therefore, also for the frequency transfer, we focus to the analysis of the two-way correction which is given by (\ref{inta}). The end-point terms given by sum of the scalar gravitational potential and centrifugal potential need to be known with uncertainty of 0.09~$\rm m^2s^{-2}$ in
order to achieve the $10^{-18}$ uncertainty of $\Delta$ (see~\cite{Wolf1} for a
detailed discussion). We focus here on the integral term in the second line of (\ref{inta}) which contains contributions of the acceleration of the fibre in the co-rotating frame and of the Coriolis and Euler acceleration. 

We define an average tangent acceleration of the fibre in the co-rotating frame as $\bar{a}_R\equiv\frac{1}{L}\int^L_0{\bf a}_R\cdot{\bf s}_l\ \D l$. This quantity should be known with uncertainty better than $\Delta\bar{a}_R=0.09\ {\rm m}^2{\rm s}^{-2}\times L^{-1}$ in order to achieve the $10^{-18}$ uncertainty in (\ref{inta}). For 1000~km long fibre it gives $9\times 10^{-8}\ {\rm ms^{-2}}$. If we consider Earth tides as the source of the fibre acceleration in the co-rotating frame then its maximal value would be around $6\times 10^{-9}\ {\rm ms^{-2}}$. 

Now we consider that the velocity of the fibre in the co-rotating frame is changed from a value ${\bf v}_R$ to a value ${\bf v}_R+\Delta{\bf v}_R$. The corresponding change of the Coriolis term in (\ref{inta}) is then given as
\begin{equation}
\int\limits_0^{L}(2\bomega\times\Delta{\bf v}_R)\cdot{\bf s}_l\ \D l\ .
\end{equation}
For this change the following upper bound can be derived 
\begin{equation}
\left|\int\limits_0^{L}(2\bomega\times\Delta{\bf v}_R)\cdot{\bf s}_l\ \D l\right|\leq 2\omega\Delta\bar{v}_{R}L
\end{equation}
where $\Delta\bar{v}_{R}$ is an average magnitude of $\Delta{\bf v}_R$ defined as
\begin{equation}
\Delta\bar{v}_{R}=\frac{1}{L}\int\limits_0^{L}||\Delta{\bf v}_R||\ \D l\ .
\end{equation}
The corresponding upper bound for change in the correction (\ref{inta}) therefore is 
$2\omega\Delta\bar{v}_{R}L/c^2$. It equals to $10^{-18}$ for $\Delta\bar{v}_{R}\approx 600\ {\rm m}^2{\rm s}^{-1}\times L^{-1}$. For an example of 1000~km long fibre it means that average uncertainty of the fibre velocity in co-rotating frame better than $0.6\ {\rm mm/s}$ is sufficient for $10^{-18}$ uncertainty in frequency transfer. If we consider
e.g. Earth tides as a source of motion of the fibre the maximal velocity of the
fibre in co-rotating frame would be around 0.05~mm/s.

Uncertainty requirement for the Earth angular velocity in the Coriolis term is not an issue since the expected value of ${\bf v}_R$ is very small.

For a change of the Euler term in (\ref{inta}) due to a variation $\Delta{\bvarepsilon}$ in the angular acceleration vector we obtain the following upper bound
\begin{equation}
\left|\int\limits_0^{L}(\Delta\bvarepsilon\times{\bf x})\cdot{\bf s}_l\ \D l\right|\leq R_m\Delta{\varepsilon}L
\end{equation}
where $R_m$ is the maximal value of $||{\bf x}||$ approximately given by the Earth radius and $\Delta{\varepsilon}=||\Delta{\bvarepsilon}||$.
The corresponding upper bound in the correction (\ref{inta}) then is $R_m\Delta{\varepsilon}L/c^2$ and it equals to $10^{-18}$ for $\Delta{\varepsilon}=1.4\times 10^{-8}\ {\rm ms}^{-2}\times L^{-1}$. For example, an uncertainty of $1.4\times 10^{-14}\ {\rm s}^{-2}$ in angular acceleration is sufficient for $10^{-18}$ uncertainty in frequency transfer for a 1000~km long fibre. But, the value of $\varepsilon$ itself usually does not exceed $5\times10^{-16}\ {\rm s}^{-2}$ as we already mentioned.

The input parameters for frequency transfer and their required uncertainties are summarized in table \ref{tabFT2}.
\begin{table}
\begin{center}
{
\small
\begin{tabular}{l c}
\hline
Parameter &Uncertainty\\
\hline
Time derivative of the fibre & \\ 
temperature (change of length& \\
and refractive index; 1-way only)& $3\times 10^{-11}$~K/s\\
Fibre velocity in co-rotating frame & 0.6~mm/s\\
Fibre acceleration& \\
in co-rotating frame& $9\times 10^{-8}\ {\rm ms}^{-2}$\\
Fibre position & $>$ Earth radius\\
Earth angular velocity & $>100$ \% (relative)\\
Earth angular acceleration & $>100$ \% (relative)\\
\hline
\end{tabular}
\caption{\label{tabFT2} Input parameters and their maximal uncertainties sufficient for $10^{-18}$ relative uncertainty in frequency transfer. The values were obtained for situations where the sensitivity of a correction to a parameter is maximized and they are calculated for 1000~km long fibre (see the text for further details and scaling of the uncertainties with the fibre length). Parameters and uncertainties related to the difference of gravitational plus centrifugal potential at the fibre endpoints are not discussed here.}
}
\end{center}
\end{table}

\section{Numerical calculation of corrections for the NPL-SYRTE and PTB-SYRTE fibre links}

In this section we evaluate the relativistic corrections for time and frequency
transfer in real fibre links connecting SYRTE (Paris, France) with NPL (Teddington,
United Kingdom) and with PTB (Braunschweig, Germany) (see
Fig.\ref{fibre_path}). These fibres are part of the REFIMEVE+
project\footnote{\url{http://www.refimeve.fr/index.php/en/}},
which aims to build a metrological fibre network in Europe.

We focus on the two-way time and frequency transfer
since the uncertainty of the leading $c^{-1}$ terms in one-way transfer coming
e.g. from an uncertainty of fibre length would lead to an inaccuracy that would
completely surpass the relativistic corrections.

For the two-way time transfer we compute the main contribution given by the
Sagnac correction~(\ref{m:twtt2}). In the two-way frequency transfer formula
(\ref{m:delta2}) the main contribution is given by the $\delta$-terms containing
the scalar potentials and velocities at the endpoints of the fibre. As we already mentioned the
evaluation of this term requires a high accuracy knowledge of the gravitational
potentials and it is a subject of a separate research.

Here we focus on the second term which comes from the time derivative of the
Sagnac term. This term is the largest contribution coming from the processes in
the fibre (not just from the state of the endpoints).   

\subsection{Discretized equations for the Sagnac correction}

For the numerical computation we suppose that we know positions of certain points of the fibre and in between these points we consider the fibre to be a straight line in Euclidean sense in the GCRS coordinates. 
We suppose that we know $N+1$ points of the fibre and we index the points by index in brackets, i.e. $(i)=1,\dots, N+1$. We have defined an initial point of the fibre $I$ and endpoint of the fibre $F$ and we suppose that the index $(i)$ is growing from $I$ to $F$ with $(i)=1$ at $I$ and $(i)=N+1$ at $F$.
We need to express the formulas for corrections in terms of coordinates of the known points and their time derivatives. We denote the spatial GCRS coordinates of the $i$-th fibre point by ${x}^a_{(i)}(t)=({x}_{(i)}(t), {y}_{(i)}(t), {z}_{(i)}(t))$. Using (\ref{m:defvs}) the Sagnac correction term can be expressed as
\begin{equation}
\int\limits_0^{L}{\bf v}\cdot{\bf s}_l\ \D l=\int\limits_0^{L}\delta_{ab}\frac{\partial{x}^a}{\partial t}\frac{\partial{x}^b}{\partial l}\D l\ .\label{SagNum}
\end{equation} 

We consider the motion of straight fibre segments as rigid body motion, i.e. the velocity $\partial{x}^a/\partial t$ changes linearly with the Euclidean length $l_E$ from one endpoint of the segment to another. In this case we obtain  
\begin{eqnarray}
&&\int\limits_0^{L}{\bf v}\cdot{\bf s}_l\ \D l=\label{vtnum}\\
&&=\frac{1}{2}\sum\limits_{i=1}^N\delta_{ab}\left(\frac{\D {x}^a_{(i)}}{\D t}+\frac{\D {x}^a_{(i+1)}}{\D t}\right)({x}^b_{(i+1)}-{x}^b_{(i)})\ .
\nonumber
\end{eqnarray}

In frequency transfer a time derivative of (\ref{SagNum}) occurs. For numerical computation we can use a formula which we obtain directly by time derivative of the right hand side of (\ref{vtnum}).  

If the effect of fibre velocity ${\bf v}_R$ in the frame co-rotating with Earth
is negligible (this velocity can appear e.g. due to the Earth tides) then we can
use the formula (\ref{SagVec}) to express the Sagnac correction in a way which
contains the Earth rotation explicitly. The Sagnac area $A$ can be expressed
using the formula~(\ref{A}). If we denote ${\bf x}_{(i)}$ a position vector connecting the center of the GCRS system with the $i$-th fibre point the Sagnac term can be expressed as
\begin{equation}
2\omega A=\sum\limits_{i=1}^N\bomega\cdot({\bf x}_{(i)}\times{\bf x}_{(i+1)})=\sum\limits_{i=1}^N\epsilon_{abc}\omega^a x_{(i)}^b x_{(i+1)}^c\ .
\label{Afnum}
\end{equation}

The time derivative needed for frequency transfer can be obtained directly by differentiating the formula (\ref{Afnum}) with respect to time.


\subsection{Fibre routing}

\begin{figure*}
\includegraphics[width=\linewidth]{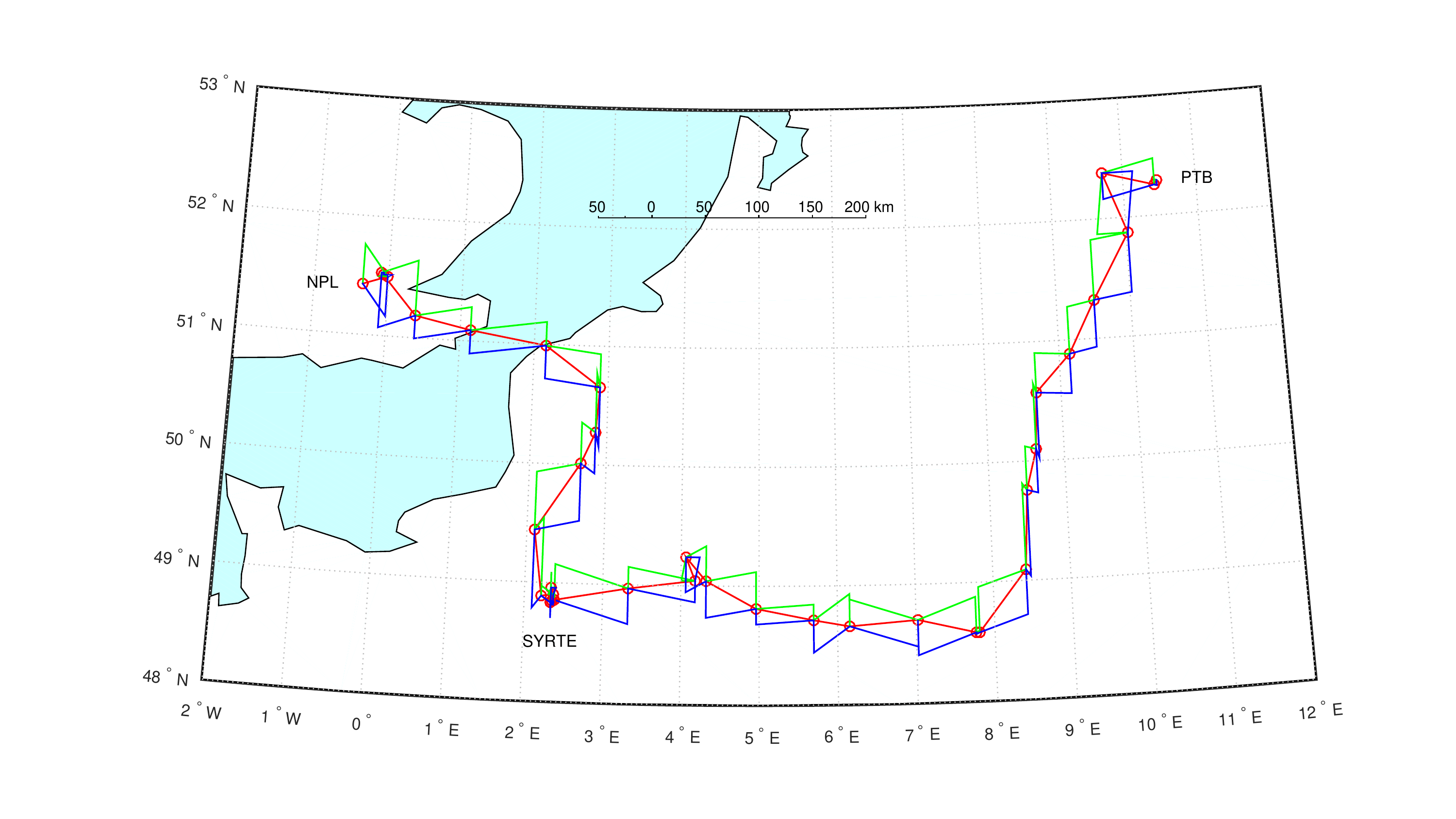}
\caption{ \label{fibre_path} Paths for the NPL-SYRTE and the PTB-SYRTE fibres.
In red: direct path (line) and positions (circles) of the shelters; in blue: maximum path, in green: minimum path.}
\end{figure*}

The path of the fibre between PTB and SYRTE, and between NPL and SYRTE, is
poorly known. However, the fibre is going through several shelters: the
positions of these shelters have been measured with GPS, and the length of the
fibre between these shelters have been measured with a laser going through the
fibre. There are 23 known shelters along the PTB-SYRTE path, and 15 along the
NPL-SYRTE path. From these data we build three different fibre paths:
\begin{enumerate}

\item the direct path: it is obtained by directly connecting the shelters with
straight lines. This path largely underestimate the total fibre length.

\item the maximum path: for this path we fix the fibre lengths between each
shelters to the measured values. An infinite number of paths can be followed by
the fibre between two given shelters, giving different values for the Sagnac
effect; we choose one that approximates the maximum of the Sagnac
effect.

\item the minimum path: as for the maximum path, the fibre lengths between each
shelters are fixed to the measured values. Then we choose a path that approximates the minimum of the Sagnac effect.

\end{enumerate}
These three paths are shown on a map in figure \ref{fibre_path}.


\subsection{Evaluation of the relativistic corrections for NPL-SYRTE and PTB-SYRTE fibre links}

\subsubsection{Two-way time transfer} 

The values obtained for the two-way time transfer Sagnac effect (\ref{m:twtt2})
can be found in table~\ref{tab_corr}. The model used is accurate to better than
1~ps, however the uncertainty in the fibre position leads to an uncertainty in
the computed correction bigger than 1~ps. We estimate the uncertainty on the
Sagnac effect due to the poorly known fibre path by calculating the Sagnac
effect for the three paths described in the previous paragraph. The differences
of the Sagnac effect between the maximum and the minimum paths are:
\begin{itemize}
\item PTB-SYRTE paths: 53~ps
\item NPL-SYRTE paths: 12~ps
\end{itemize}

There are also variations of the Sagnac term due to Earth tides
and due to changes in the angular velocity vector of the Earth. All these variations, however,
lead to corrections of order or less than 5~fs. Tides also imply a residual
velocity with respect to the co-rotating frame, which lead to a correction
which is much less than 1~ps.

\begin{table}
\begin{center}
\begin{tabular}{|l||c|c|}
\hline
fibre link & Length/km & Correction/ps\\
\hline\hline
PTB-SYRTE & 1401 & 3976 $\pm$ 27\\
\hline
NPL-SYRTE & 813  & 1214 $\pm$ 6\\
\hline
\end{tabular}
\caption{\label{tab_corr} Relativistic corrections for two-way time transfer in case of the PTB-SYRTE and NPL-SYRTE fibre links. The only contribution relevant for the 1~ps accuracy comes from the Sagnac term. The $\pm$ values indicate the estimated maxima and minima of the correction as described in the text.}
\end{center}
\end{table}

\subsubsection{Two-way frequency transfer} In time derivative of the Sagnac correction in (\ref{m:delta2})
there are contributions coming from the Earth tides and from the variation of the angular velocity vector of the Earth. All these contributions are varying and have amplitudes of order $10^{-19}$ or less,
which corresponds to a variation of around 5~fs over 12 hours in the time
transfer.

These terms are not studied in detail here as they are below the accuracy goal.
However it would be interesting to model them accurately as they are just one
order of magnitude below the limit of the actual clock stability. By integrating the
clock frequency comparison for a long time it could be detectable.


\section{Conclusion}
In this work we presented a systematic relativistic description of propagation of a signal in optical fibres. We derived a general differential equation governing the signal propagation and we investigated its solution up to terms of order $c^{-3}$. This is sufficient for accuracy of 1 ps for time transfer formulas and for relative accuracy of $10^{-18}$ for frequency transfer formulas. This accuracy corresponds to the requirements of nowadays and near future optical clocks applications. 

Formulas for both one-way and two-way time and frequency transfer were derived. The largest relativistic correction for one-way and two-way time transfer is the so called Sagnac effect which strongly depends on the fibre positioning on Earth surface and its magnitude usually does not exceed order of nanoseconds. The next significant correction for one-way transfer takes gravitational field into account and depends on the gravity potential along the fibre. This correction usually does not exceed order of picoseconds. 

For frequency transfer there is the well known gravitational red shift which depends only on state of the endpoints of the fibre. Moreover there are other effects which depend on the processes in the fibre itself. For two-way transfer the largest effects are due to change of the fibre position with time and due to thermal variations. Contribution of these effects was estimated to be smaller than $10^{-18}$. 

\section*{Acknowledgment} The research has received funding from the European
Metrology Research Programme (EMRP). The EMRP is jointly funded by the EMRP
participating countries within Euramet and the European Union. The work have
been done within an EMRP project International Time Scales with Optical Clocks and related Research Mobility Grant.
The authors would like to thank Paul-Eric Pottie and Fabio Stefani for
the data on optical fibres and fruitful discussions, and Isabelle Panet
and Christian Bizouard for their help and advices.

\appendix
\section{Theory}
\label{App}

\subsection{Constrained relativistic propagation equation of signal in a medium}

We use the following conventions. Small Greek indices go from 0 to 3, small Latin indices go from 1 to 3 and metric signature is $-+++$.

First we will derive a condition governing signal propagation in optical fibre. If we denote $U^\alpha$ the four-velocity field of the fibre motion and $k^\alpha$ a tangent vector to a trajectory of the signal propagating in the fibre the condition (\ref{m:neff}) can be written in terms of components of vector $k^\alpha$ in the rest frame of the observer $U^\alpha$ as (we drop the index eff)
\begin{equation}
c\frac{\sqrt{\delta_{ab}k^ak^b}}{k^0}=\frac{c}{n}.\label{velo_cond}
\end{equation}
The terms containing the components of vector $k^\alpha$ can be written in covariant form as follows
\begin{eqnarray}
k^0&=&-U_\alpha k^\alpha\label{temporal}\\
\delta_{ab}k^ak^b&=&(g_{\alpha\beta}+U_\alpha U_\beta)k^\alpha k^\beta\label{spatial}
\end{eqnarray}
where $g_{\alpha\beta}$ is spacetime metric and the expression for $\delta_{ab}k^ak^b$ was obtained as a  square of orthogonal projection of $k^\alpha$ into the normal space of $U^\alpha$ which is given as $(\delta^\alpha_\beta+U^\alpha U_\beta)k^\beta$.
Inserting (\ref{temporal}) and (\ref{spatial}) into (\ref{velo_cond}) we obtain the following condition for the vector $k^\alpha$
\begin{equation}
n\sqrt{(g_{\alpha\beta}+U_\alpha U_\beta)k^\alpha k^\beta}=-U_\alpha k^\alpha\ .
\label{prop_cond}
\end{equation}
This condition can be also written as $\gamma_{\alpha\beta}k^\alpha k^\beta=0$ with $\gamma_{\alpha\beta}$ being defined as
\begin{equation}
\gamma_{\alpha\beta}\equiv g_{\alpha\beta}+\left(1-\frac{1}{n^2}\right)U_\alpha U_\beta.
\label{optMet}
\end{equation}
It means that the signal propagates along null lines of a modified metric $\gamma_{\alpha\beta}$.
Besides exchanging a refractive index for the effective refractive index this is exactly the metric introduced by Gordon in  \cite{Gordon}.

Now we are going to derive an equation for signal propagation in optical fibre based on the condition (\ref{prop_cond}). 

We consider a spacetime with time coordinate $t$. The fibre is considered to be a one dimensional object, i.e. its trajectory is 2D surface in spacetime. Fibre is parametrised by two parameters - the coordinate time $t$ and a parameter $\lambda$ which has a unique value for a given fibre element. The parameter $\lambda$ has a range $\lambda\in[\lambda_I,\lambda_F]$ and is growing from an initial point of the fibre which we denote $I$ to the final point of the fibre which we denote $F$. If $x^\alpha$ are coordinates on the spacetime the fibre worldsurface is given parametrically as
\begin{equation}
x^\alpha=x^\alpha(t,\lambda)\ .
\end{equation}
For a fixed value of $\lambda$ this equation gives a trajectory of a corresponding fibre element. Therefore the four-velocity of the fibre $U^\alpha$ is proportional to a partial derivative of $x^\alpha(t,\lambda)$ with respect to time. We have
\begin{equation}
U^\alpha =U^\beta\partial_\beta t\ \frac{\partial x^\alpha}{\partial t}
\label{U}
\end{equation}
where the proportionality factor can be expressed using a normalisation condition $U^\alpha U_\alpha =-1$ as
\begin{equation}
 U^\alpha\partial_\alpha t =\left(-g_{\alpha\beta}\frac{\partial x^\alpha}{\partial t}\frac{\partial x^\beta}{\partial t}\right)^{-\frac{1}{2}}.
\label{Utot}
\end{equation}

The parameters $t, \lambda$ can be seen as coordinates on the worldsurface of the fibre. The signal trajectory can be therefore described by functions $t(\sigma), \lambda(\sigma)$ where $\sigma$ is some parametrisation of the signal trajectory. In coordinates $x^\alpha$ the trajectory reads $x^\alpha(t(\sigma),\lambda(\sigma))$. It will be convenient to choose the parameter $\sigma=\pm\lambda$ where $+$ is for propagation from $I$ to $F$, i.e. with increasing $\lambda$, and $-$ is for propagation from $F$ to $I$, i.e. with decreasing $\lambda$. In this case the tangent to the trajectory can be expressed as
\begin{equation}
k^\alpha=\frac{\D x^\alpha}{\D\sigma}=\pm\frac{\partial x^\alpha}{\partial t}\frac{\D t}{\D\lambda}\pm\frac{\partial x^\alpha}{\partial \lambda}.
\label{tangent}
\end{equation}

Inserting (\ref{tangent}) to (\ref{prop_cond}) with use of (\ref{U}) and with use of the fact that projection of $\partial x^\alpha/\partial t$ to normal space of $U^\alpha$ is zero we obtain the following equation
\begin{equation}
\frac{\D t}{\D \lambda}=U^\alpha\partial_\alpha t\left(\pm n\frac{\partial l_t}{\partial \lambda}+U_\beta\frac{\partial x^\beta}{\partial\lambda}\right)
\label{propeq2}
\end{equation}
where we defined a function
\begin{equation} 
l_t(t,\lambda)=\int\limits_{\lambda_I}^{\lambda}\sqrt{(g_{\alpha\beta}+U_\alpha U_\beta)\frac{\partial x^\alpha}{\partial\lambda}\frac{\partial x^\beta}{\partial\lambda}}\ \D\lambda
\label{lt}
\end{equation}
which is a rest length of the fibre at a time $t$ between elements $\lambda_I$ and $\lambda$.

The equation (\ref{propeq2}) is a differential equation for the function
$t(\lambda)$ which describes the trajectory of the signal in the fibre in terms
of coordinate time $t$ at which the signal reaches an element $\lambda$ of the
fibre. The sign $+$ in the equation relates to the signal propagation from $I$
to $F$, i.e. along the orientation of the $\lambda$ coordinate and the sign $-$
relates to propagation from $F$ to $I$. 

\subsection{Propagation equation for optical fibre on Earth surface}\label{App2}

Now we are going to express the right hand side of the equation (\ref{propeq2}), for a particular case of a fibre located on Earth surface. We can do it using the formulas (\ref{U}) and (\ref{Utot}) where $x^\alpha$ are chosen to be the GCRS coordinates $(ct,x,y,z)$ with metric components $g_{\alpha\beta}$ given by (\ref{m:g00}), (\ref{m:g0i}) and (\ref{m:gij}). The derivatives $\partial x^\alpha/\partial t$ and $\partial x^\alpha/\partial\lambda$ which occur in (\ref{U}), (\ref{Utot}) and (\ref{propeq2}) can be expressed in terms of the velocity vector field and tangent vector field of the fibre defined by (\ref{m:defvs}). 

Moreover an Euclidean vector corresponding to the vector gravitational potential $w_i$ can be defined as 
\begin{equation}
{\bf w}=w_i\delta ^{ij}{\bf e}_j\ .
\end{equation}

Expanding the right hand side of the equation (\ref{propeq2}) in powers of $c^{-1}$ we obtain
\begin{eqnarray}
\frac{\D t_\pm}{\D\lambda}&=&\pm\frac{n}{c}\frac{\partial l_t}{\partial\lambda}+\frac{1}{c^2}{\bf v}\cdot {\bf s}_\lambda \pm\frac{1}{c^3}\!\left(w+v^2/2\right)n\frac{\partial l_t}{\partial\lambda}\label{propeq3}\\
&+&\frac{1}{c^4}\left((4w+v^2){\bf v}\cdot{\bf s}_\lambda -4{\bf w}\cdot {\bf s}_\lambda\right)+O(c^{-5})
\nonumber
\end{eqnarray}
where we denoted $v^2={\bf v}\cdot{\bf v}$ and we added the $\pm$ sign also to the left side of the equation to distinguish the solutions for propagation in different directions. 

The first term on the right hand side of (\ref{propeq3}) corresponds to the Newtonian limit. The second term corresponds to the Sagnac correction. The third term corresponds to a part of the Shapiro correction. Another part would appear if we would use an Euclidean length of the fibre in the GCRS coordinates as the parameter $\lambda$ (see the discussion below). The fourth term contains the geodetic and Lense-Thirring effect.

The parameter $\lambda$ in (\ref{propeq3}) is an arbitrary parameter satisfying the condition that its value is unique (not changing in time) for a given fibre element. However, some natural choices can be done which are convenient for practical purposes. One possibility for such a parameter is the rest length $l_t$ itself at a specific time $t_1$. The fibre can expand e.g. due to temperature changes and therefore $l_t$ must be fixed at certain time to fulfill the uniqueness condition. We denote this parameter $l$, i.e. $l(\lambda)=l_t(t_1,\lambda)$. If we consider the thermal expansion we have
\begin{equation}
\frac{\partial l_t}{\partial l}=1+\alpha(T(t,l)-T(t_1,l))
\label{Texpansion}
\end{equation}
where $T(t,l)$ is temperature of the fibre at given time and position and $\alpha$ is its linear thermal expansion coefficient.

Another choice of the parameter could be an Euclidean length of the fibre at a time $t_1$. The Euclidean length at a time $t$ between the fibre elements $\lambda_I$ and $\lambda$ is defined as
\begin{equation}
l_{Et}(t,\lambda)=\int\limits^\lambda_{\lambda_I}\sqrt{\delta_{ab}\frac{\partial x^a}{\partial\lambda}\frac{\partial x^b}{\partial\lambda}}\ \D\lambda\ .
\end{equation}
We denote the Euclidean parameter of the fibre $l_E$, i.e. $l_E(\lambda)=l_{Et}(t_1,\lambda)$. We obtain
\begin{equation}
\frac{\partial l}{\partial l_E}=1+\frac{1}{c^2}\left(w+\frac{({\bf v}\cdot{\bf s})^2}{2}\right)+O(c^{-4}) 
\label{EuclidRest}
\end{equation}
with ${\bf s}$ being a unit tangent to the fibre, i.e. ${\bf s}\cdot{\bf s}=1$. 

Choosing $l_E$ as the parameter of the fibre then leads to the terms in (\ref{propeq3}) given by ${\partial l_t}/{\partial l_E}=({\partial l_t}/{\partial l}).({\partial l}/{\partial l_E})$ with the derivatives on the right hand side given by (\ref{Texpansion}) and (\ref{EuclidRest}).


\subsection{Solving the equation for signal propagation \label{2.2}}

Now we are going to solve the differential equation (\ref{propeq3}). We denote the right hand side of (\ref{propeq3}) by $\Omega _{\pm}(t,\lambda)$ with $+$ sign referring to propagation from $I$ to $F$ and $-$ sign referring to propagation in opposite direction. Then the equation (\ref{propeq3}) can be written as

\begin{equation}
\frac{\D t_\pm}{\D \lambda}=\Omega _{\pm}(t,\lambda)\ .
\label{propEq}
\end{equation}

If $\Omega_\pm$ does not depend on $t$ we can integrate the equation directly. 
However, $\Omega _{\pm}$ depends on time $t$, because of e.g. thermal
expansion, Earth tides or variations of the instantaneous rotation vector of the
Earth. Nevertheless, these time dependencies are very slow, and we will see that
the value of $\Omega _{\pm}$ does not change significantly during the
propagation time of the signal from one end to the other end of the fibre, which
is around 5~ms for a 1000~km fibre with index $n=1.5$.

We consider a signal emitted from observer $I$ at time $t_0$, reflected from observer $F$ at time $t_1$ and received back by observer $I$ at time $t_2$ as depicted in figure \ref{fig:2way}. We use $l$ as the parameter of the fibre which was defined as $l(\lambda)=l_t(t_1, \lambda)$. The range of the parameter is $l\in[0,L]$ with $L=l(\lambda_F)$ being the total rest length of the fibre at time $t_1$.

Formal solution of the equation (\ref{propEq}) can be written in a form
\begin{equation}
t_\pm(l)=t_1+\int\limits_{L}^l\Omega_\pm(t_\pm(l), l)\ \D l\ .
\label{genSol}
\end{equation}
The formula (\ref{genSol}) can be further processed by iterations. In the first iteration we set $t_\pm(l)=t_1$ on the right-hand-side of (\ref{genSol}). Therefore we obtain 
\begin{equation}
t^{[1]}_\pm(l)=t_1+\int\limits_{L}^{l}\Omega _{\pm}(t_1,l)\ \D l
\end{equation}
where we marked the iteration number in square bracket. The second iteration then gives
\begin{equation}
t^{[2]}_\pm(l)=t_1+\int\limits_{L}^{l}\Omega _{\pm}\left(t_1+\int\limits_{L}^{l}\Omega _{\pm}(t_1,l^\prime)\ \D l^\prime,\ l\right)\D l.
\label{oTTp}
\end{equation}
The order of deviation of the exact solution from the second iteration is given by
\begin{equation}
t_\pm(l)=t^{[2]}_\pm(l)+\Or\left[\left(\!\frac{\partial\Omega_\pm}{\partial t}\!\right)^{\!\!2}\!\!\delta l^2\delta t(l)\right]
\end{equation}
where $\delta l=l-L$ and $\delta t(l)$ is propagation time between $l$ and $L$. This deviation turns out to be negligible for our level of accuracy, e.g. for 1000~km long fibre and for effects caused by temperature variations it gives $10^{-28}\ {\rm s}$. Therefore we use the formula (\ref{oTTp}) for $t_\pm(l)$.

Now we expand $\Omega_{\pm}(t,l)$ in (\ref{oTTp}) in time around $t_1$. We obtain
\begin{equation}
t_\pm(l)=t_1+t_\pm^{(0)}(l)+t_\pm^{(1)}(l)+\Or\left[\frac{\partial^2\Omega_\pm}{\partial t^2}\delta l\delta t(l)^2\right]
\label{eq:OmExp}
\end{equation}
where
\begin{eqnarray}
t_\pm^{(0)}(l)&=&\int\limits_{L}^l\Omega_\pm(t_1,l)\D l\label{oTTp2}\\
t_\pm^{(1)}(l)&=&\int\limits_{L}^l\frac{\partial\Omega_\pm}{\partial t}(t_1, l)\left[\int\limits_{L}^l\Omega_\pm(t_1,l^\prime)\D l^\prime\right]\D l\ .\label{oTT3}
\end{eqnarray}

The term of (\ref{oTT3}) in square bracket can be expressed using the Newtonian limit of $\Omega_\pm$ given by the $c^{-1}$ term of (\ref{propeq3}) where we consider the refractive index to be a constant which we denote $n_0$. Thus we get 
\begin{equation}
t^{(1)}_\pm(l)=\pm\frac{n_0}{c}\int\limits_{L}^l\frac{\partial\Omega_\pm}{\partial t}(t_1, l)(l-L)\D l\ .
\label{oTT}
\end{equation}

Now we can check the magnitude of the contribution (\ref{oTT}). For our estimation we use the Newtonian limit of $\Omega_\pm$ to calculate $\partial\Omega_\pm/\partial t$ again. We obtain 
\begin{equation}
\frac{\partial\Omega_\pm}{\partial t}(t_1,l)\approx\pm\frac{\partial}{\partial t}\!\left(\frac{n}{c}\frac{\partial l_t}{\partial l}\right)\!=
\pm\frac{1}{c}\!\left(\frac{\partial n}{\partial t}+n\alpha\frac{\partial T}{\partial t}\right)
\label{dOdt}
\end{equation}
where we considered that the expansion of the fibre is caused by a change of temperature and we used the formula (\ref{Texpansion}). The order of (\ref{oTT}) then is
\begin{equation}
t_\pm^{(1)}(l)\sim\frac{1}{2}\frac{n_0}{c^2}\left(\frac{\partial n}{\partial t}+n\alpha\frac{\partial T}{\partial t}\right)(l-L)^2\ .
\label{t1order}
\end{equation}
For our estimation we consider a change of $n$ due to changing temperature, i.e. $\partial n/\partial t =(\partial n/\partial T).(\partial T/\partial t)$. Based on the values given by (\ref{alpha}) and (\ref{dTdt}) we estimate:
\begin{equation}
t_\pm^{(1)} \sim 2 \times 10^{-11}{\rm s}^{-1} \times n_0 \left(\frac{L}{c} \right)^2 \sim 0.3 \
\textrm{fs} \nonumber
\end{equation}
for a fibre with $L \sim 1000$~km and $n_0 \sim 1.5$. Therefore, a maximum
variation of 0.6~fs occurs in 12~h, corresponding to a change in the relative
frequency comparison of $\sim 1.5 \times 10^{-20}$, below the required
accuracy.

In the following we will not consider the contribution $t_\pm^{(1)}$. 

Finally we get the solution to the required accuracy
\begin{equation}
t_\pm(l)=t_1+\int\limits_L^l\Omega_\pm(t_1,l)\D l\ .\label{oTTp3}\label{oTTm3}
\end{equation}

\subsection{Time transfer}

In time transfer the coordinate times of signal propagation between the fibre endpoints are needed. 
We denote $\Delta t_+$ the coordinate time of signal propagation from $I$ to $F$ and $\Delta t_-$ the coordinate time of signal propagation from $F$ to $I$, i.e. 
\begin{eqnarray}
\Delta t_+ &=& t_1-t_0 = t_1-t_+(0)\ ,\label{dtp}\\ 
\Delta t_- &=& t_2-t_1 = t_-(0)-t_1\ .\label{dtm}
\end{eqnarray}
Using the formula (\ref{oTTp3}) we obtain 
\begin{equation}
\Delta t_\pm=\pm\int\limits_0^L\Omega_\pm(t_1,l)\ \D l\ .
\label{Dtpm}
\end{equation}

Inserting the $\Omega_\pm$ given by (\ref{propeq3}) into (\ref{Dtpm}) up to the $c^{-3}$ order we obtain the formula (\ref{m:dtp1}).



\subsection{Frequency transfer}

For one-way frequency transfer we start with the formula (\ref{m:fratio}) describing the ratio of proper frequencies of a signal as observed during its emission at one endpoint of the fibre and reception at the opposite endpoint. 

The derivatives $\D t_e/\D\tau_e$ and $\D t_r/\D\tau_r$ in (\ref{m:fratio}) are discussed in the main part of the text. 

The derivative $\D t_r/\D t_e$ in (\ref{m:fratio}) can be evaluated using the formula~(\ref{oTTp3}). For propagation from $I$ to $F$ we have $\D t_{r+}/\D t_e=({\partial t_+(0)}/{\partial t_1})^{-1}$ and for propagation from $F$ to $I$ we have ${\D t_{r-}}/{\D t_e} = {\partial t_-(0)}/{\partial t_1}$. Therefore we get
\begin{equation}
\frac{\D t_{r\pm}}{\D t_e} =1 \pm\int\limits_0^{L}\frac{\partial\Omega_{\pm}}{\partial t}(t_1,l)\ \D l
\label{dtdti}
\end{equation}
where we used $(1-\varepsilon)^{-1}\approx 1+\varepsilon$ for the $+$ direction and therefore we neglected terms of order $\varepsilon^2$, i.e. of order $(\partial\Omega_\pm/\partial t)^2L^2\sim 10^{-38}{\rm m}^{-2}\times L^2$ in a case of thermal variations of $\Omega_\pm$.

In the formula (\ref{dtdti}) we also neglected the terms coming from the higher order variations of $\Omega_{\pm}$. In particular, we have seen that the
first order term of expansion~(\ref{eq:OmExp}) given by (\ref{t1order}) leads to a correction in the frequency transfer:
\begin{equation}
\frac{\D t^{(1)}_{r\pm}}{\D t_e} \sim 9\times 10^{-16}{\rm s}^{-2} \times n_0 \left(\frac{L}{c} \right)^2
\sim 1.5 \times 10^{-20} \nonumber
\end{equation}
for a fibre with $L \sim 1000$~km and $n_0 \sim 1.5$.

Inserting the $\Omega_\pm$ given by (\ref{propeq3}) into (\ref{dtdti}) up to the $c^{-2}$ order and with use of (\ref{dOdt}) we obtain the formula (\ref{m:dtdti2}).

The two-way frequency transfer is described by the formula (\ref{m:2wayf}). This
formula can be understood as definition of the correction $\Delta$ which needs
to be computed. The $\Delta$ can be expressed from (\ref{m:2wayf}) and further
processed using the formula (\ref{m:fratio}) for the frequency ratios with $\D
t_r/\D t_e$ given by (\ref{dtdti}) and $\D t/\D \tau$ calculated from the
metric.

We write $\D t/\D \tau=1+\delta$ with $\delta\ll 1$ (on the Earth surface
$\delta \sim 10^{-9}$).
Therefore we neglect terms where $\delta$ multiplies the integrals of
$\partial\Omega_\pm/\partial t$ which, for the case of thermal variations of $\Omega_\pm$, are of order $L (\partial\Omega_\pm/\partial t) \delta \sim 10^{-28}{\rm m}^{-1} \times L$, far below the required
accuracy. We also neglect the $\delta^2$ terms since their contribution together with $c^{-4}$ terms of $\delta$ is of order below $10^{-24}$ for altitude difference of the fibre endpoints below $10$~km at the Earth surface. Using these approximations we obtain the following formula for the computed correction
\begin{eqnarray}
\Delta &=&(\delta_{I0}+\delta_{I2})/2-\delta_{F1}\nonumber\\
&+&\frac{1}{2}\int\limits_0^{L}\left[\frac{\partial\Omega_+}{\partial t}(t_1, l)+\frac{\partial\Omega_-}{\partial t}(t_1, l)\right]\D l\label{delta}
\end{eqnarray}
where the indices $I0$, $I2$ mean that the quantity is evaluated at the position of the observer $I$ at the time $t_0$ or $t_2$ respectively and similarly the index $F1$ means the quantity is evaluated at the position of the observer $F$ and time $t_1$. In (\ref{delta}) we can approximate 
\begin{equation}
(\delta_{I0}+\delta_{I2})/2\approx\delta_{I1}
\label{dapp}
\end{equation}
since the non-linearity of $\delta_I(t)$ causes deviations which are far below the required accuracy.

Inserting (\ref{dapp}) and the $\Omega_\pm$ given by (\ref{propeq3}) up to the $c^{-3}$ order into (\ref{delta}) we obtain the formula (\ref{m:delta2}).


\end{document}